\documentclass[twocolumn]{jpsj2} 
\newcommand{\bra}{\left\langle}
\newcommand{\ket}{\right\rangle}

\title{Fluctuation theorem applied to {\it Dictyostelium discoideum} system 
}

\author{Kumiko \textsc{Hayashi}$^{1}$ and Hiroaki \textsc{Takagi}$^{2}$
\thanks{Present adress: Department of Physics, Nara Medical university, 
Nara, Japan.
E-mail address: takagi@naramed-u.ac.jp}}

\inst{$^{1}$ Department of Physics, Waseda University, Tokyo 169-8555, 
Japan
  \\
$^{2}$ Graduate School of Frontier Biosciences,  Osaka University, 
Osaka 565-0871, Japan}

\kword{non-equilibrium statistical mechanics, fluctuation theorem, 
cell motility, electrotaxis, Dictyostelium}

\begin{document}
\maketitle

Directional cell movement is a fundamental phenomenon exhibited by 
many biological processes.  It has been known that an electric field 
exists on the surfaces of the tissues of organisms  and that it acts 
as a directional cue for the type of the cell migration known as 
electrotaxis.  Electrotaxis is thought to play important roles in 
various physiological processes including embryogenesis and wound 
healing, and the underlying molecular mechanisms of electrotaxis are 
now extensively studied \cite{nature}.

\begin{figure}[h]
\begin{center}
\includegraphics[width=2.7cm]{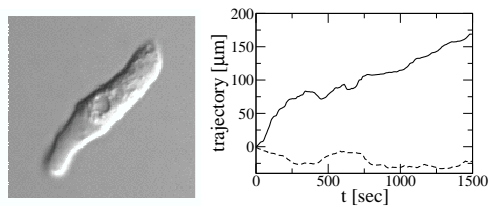} 
\includegraphics[width=3.7cm]{fig2.eps}  
\end{center}
\caption{(Left) {\it Dictyostelium discoideum}. The lower edge, with the 
smooth region,  is the front side of the cell. Its length is of the order 
of 10 $\mu$m. (Right)  Experimental trajectories, $x(t)-x(0)$ (solid line) 
and $y(t)-y(0)$ (dotted line), are plotted as functions of time in the 
case $E=10$ V/cm.
}
\label{fig:monyo}
\end{figure}

In order to elucidate the mechanisms of the electrotactic responses of 
cells, the cellular slime mold {\it Dictyostelium discoideum} (see Fig. 
\ref{fig:monyo} (Left)) is a suitable organism to study, because of its 
high motility and strong electrotactic response.   With well-established 
genetic engineering techniques and advanced microscopic techniques, 
the input-output relationship in the electrotactic response of 
{\it Dictyostelium} cells has been investigated to elucidate the 
stochastic processes involved in the signaling systems responsible for 
cell motility and their regulations \cite{sato1}.

In this paper, we analyze the electrotactic movement of 
{\it Dictyostelium discoideum} from the viewpoint of non-equilibrium 
statistical mechanics.  Because we can observe fluctuating behavior of 
cellular trajectories,  we analyze the probability distribution of the 
trajectories  with the aid of the fluctuation theorem.  Recently, the 
validity of the fluctuation theorem was verified in a colloidal system 
\cite{ft}, and it has also been applied to granular systems \cite{gft1},  
turbulent systems \cite{gft3}  and chemical oscillatory waves \cite{gft2} 
to investigate some of their statistical properties that are not yet 
completely understood.  Noting that the fluctuation theorem is 
potentially applicable to cellular electrotaxis, here we employ it to 
help us obtain a phenomenological model of this biological system. 

\begin{figure}[h]
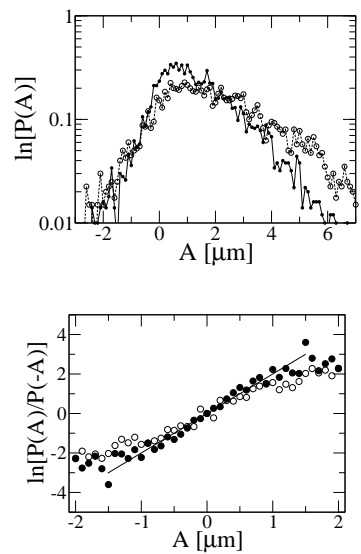

\begin{center}
\includegraphics[width=4.6cm]{fig3.eps} \\
\vspace{6mm}
\includegraphics[width=4.25cm]{fig4.eps}
\end{center}
\caption{(Top) The probability distribution, ${\rm P}(A)$, in the case 
$E=10$ V/cm with $\tau=10$ sec (filled circles) and $\tau=15$ sec (open 
circles). $4000$ samples of $A$ were obtained from $58$ independent 
trajectories.  (Bottom) The quantity $\ln[{\rm P}(A)/{\rm P}(-A)]$ as 
a function of $A$ in the case  $\tau=10$ sec (filled circles) and the case 
$\tau=15$ sec (open circles). The slope of the straight line is equal to 
$2$ $\mu{\rm m}^{-1}$.   
 } 
\label{fig:prob}
\end{figure}

{\it System}. In this study, {\it Dictyostelium discoideum}, Ax2 cells 
(wild type) were starved for up to $4$ hours, with a pulse of $100$ nM 
cAMP applied every $6$ min. The cell suspension was injected into the 
chamber for electrotactic assay, and the cells were allowed to spread 
over the coverslip for $20$ min at $T=294$ K.  Direct current was applied 
to the chamber as illustrated in the paper of Sato {\it et al.} \cite{sato1}  

The cells in the chamber were observed with a microscope capable of 
producing differential interference contrast optics. Data acquisition 
started 5 min after the electric field was first applied, and the electric 
field remained at a constant strength, $E$, throughout experiments.  Under 
these conditions, we considered the system to be in a steady state.  To 
analyze the motile activities of the cells under the electric field, cell 
images were processed automatically with a time resolution of $5$ sec and 
converted into binary images by selecting an optimal value of the 
brightness threshold. In this way, the trajectory of the center position 
of the observed cellular region, $(x(t),y(t))$, was determined. Because the 
gradient of the electric field is non-zero only along the $x$ direction, 
we particularly investigate the trajectories in the $x$ direction for the  
case that $E=10$ V/cm, which is sufficiently large for the response 
saturation of the cell \cite{sato1}.  

{\it Fluctuation theorem}. 
When the electric field is turned on, {\it Dictyostelium} cells begin to 
migrate toward the cathode (the $+x$ direction). In Fig. \ref{fig:monyo} 
(Right),  we plot example trajectories, $x(t)-x(0)$ and  $y(t)-y(0)$, as 
functions of time in the case $E=10$ V/cm.  The fluctuating behavior of 
the cell movement was observed.  Then, using these fluctuating trajectories 
and setting $A\equiv x(\tau)-x(0)$, we investigate the probability density 
for the realization of the value $A$. In Fig. \ref{fig:prob} (Top), we plot 
the probability density, ${\rm P}(A)$, in the case $E=10$ V/cm with  $\tau=10$ 
sec and $\tau=15$ sec. We find that the graphs have  exponential tails.  

In a colloidal system under a non-equilibrium condition, an entropy 
production, $\Sigma$, of the system was measured experimentally \cite{ft}, 
and the probability distribution, ${\rm P}(\Sigma)$, was analyzed.  This 
${\rm P}(\Sigma)$ also has exponential tails, and it is known that the 
entropy production fluctuation theorem $\ln[{\rm P}(\Sigma)/{\rm P}
(-\Sigma)]=\Sigma$ holds in the system. Because the forms of the graphs 
in Fig. \ref{fig:prob} (Top) are similar to those observed in the colloidal 
system,  we plot $\ln [{\rm P}(A)/{\rm P}(-A)]$ as a function of $A$ in 
Fig. \ref{fig:prob} (Bottom).  In Fig. \ref{fig:prob} (Bottom),  we obtain 
the linear slopes in the cases $\tau=10$ sec and $\tau=15$ sec, and they 
differ slightly. We thus conclude that the value of the slopes is almost 
independent of $\tau$ at least to the extent of this rough treatment. From 
the fitting of the data to a linear function, we obtain the value of the 
slopes $2$ $\mu{\rm m}^{-1}$.

In order to understand a physical meaning of the value $2$ $\mu{\rm m}^{-1}$,  
we employ a model simplified enough, noting that $\tau$ is less than the 
persistence time of the cell movement.  We consider that a description of 
the cell movement is roughly described by the Langevin equation  
\begin{eqnarray}
\Gamma\frac{{\rm d} x}{{\rm d} t}&=&F+\Xi(t), 
\nonumber \\ 
\bra \Xi(t)\Xi(t')\ket &=& 2\alpha\Gamma\delta(t-t'), 
\label{model}
\end{eqnarray}
where $\Gamma$ is the effective friction coefficient, $F$ is the apparent 
driving force,  and $\alpha$ is a parameter that characterizes the 
fluctuations of the cell movement.  

A physical meaning of the value $2$ $\mu{\rm m}^{-1}$, which is almost  
independent of $\tau$,  is understood by one expression of the 
fluctuation theorem.  For the model (\ref{model}), we can derive the 
relation 
\begin{equation}
\ln \left[ \frac{{\rm P}(A)}{{\rm P}(-A)}\right]
=\frac{F}{\alpha}A.  
\label{ft}
\end{equation}
The relation (\ref{ft}) means that  we can infer the value of $F/\alpha$  
by computing ${\rm P}(A)$, ${\rm P}(-A)$ and $A$ from the experimental 
data \cite{sato1}, and that the value of the slope $2$ $\mu$m$^{-1}$ 
in Fig. \ref{fig:prob} (Bottom)  corresponds to $F/\alpha$.

{\it Discussion}. 
Because the traction force of a {\it Dictyostelium} cell has been measured 
experimentally and found that it is of the order of $1$ nN \cite{uchida}, 
$F$ is also considered to be of the order of $1$ nN.  Then, using  
$F/\alpha=2$ $\mu$m$^{-1}$, $\alpha$ is estimated to be $0.5$ nN$\mu$m 
(= $10^{5}$ $k_{\rm B}T$ \cite{eff}).  
Using the relations $\Gamma\bra v\ket_{E=10}=F$ and $\bra v\ket_{E=10}=0.1$ 
$\mu$m/sec, $\Gamma$ is also estimated to be $10$ g/sec.  Here, $v(t)$ 
is the velocity of a cell defined by $v(t)\equiv (x(t+\Delta t)-x(t))/
\Delta t$ where $\Delta t= 5$ sec, and $\bra \ \ket_{E}$ denotes the 
steady state average.  

In this paper, we obtained the rough estimation of the parameters $\Gamma$, 
$F$ and $\alpha$ that characterize the cell movement under electrotaxis by 
using the fluctuation theorem  (\ref{ft}), which is derived for the simple 
model (\ref{model}).  To confirm the validity of the characterization, the 
above parameter values should be compared with those of other organisms, 
e.g. {\it Amoeba proteus} \cite{miyoshi}. 
Also in future, the presented method should be extended to satisfy more 
precise descriptions of cell movement \cite{takagi,Flyvbjerg}, noting that 
the fluctuation theorem can be derived for more generic stochastic processes 
\cite{kurchan}.

In general, theories of non-equilibrium statistical mechanics, such as the 
fluctuation theorem and the Jarzynski equality,  have been applied to 
small non-equilibrium systems,  such as colloidal systems, 
RNA systems and protein systems.  However, as there are few studies in which 
these theories are applied to cell systems, we hope that the present study 
will help lead to exploration in this new direction. 

\acknowledgements 

The authors acknowledge M. J. Sato and M. Ueda for useful discussions of 
all of the issues considered  in this paper and Y. Iwadate for informing 
them of the paper \cite{uchida}.  One of the authors (K. H.) appreciates the 
members of the Yanagida group for their hospitality during her visit to 
Osaka University and the members of the Takano group for providing her 
a research environment. This work was supported by grants form JSPS 
Research Fellowships for Young Scientists and Leading Project (Bio Nano 
Process),  MEXT, Japan.


\begin{thebibliography}{99} 


\bibitem{nature}
M. Zhao et al.: Nature {\bf 442} (2006) 457.  

\bibitem{sato1}
M. J. Sato, M. Ueda, H. Takagi, T. M. Watanabe, T. Yanagida, and M. Ueda: 
Biosystems {\bf 88(3)} (2007) 261. 

\bibitem{ft}
G. M. Wang, E. M. Sevick, E. Mittag, D. J. Searles, and D. J. Evans: Phys. 
Rev. Lett. {\bf 89} (2002) 050601. 

\bibitem{gft1}
K. Feitosa and N. Menon: Phys. Rev. Lett. {\bf 92} (2004) 164301. 

\bibitem{gft3}
S. Ciliberto, N. Garnier, S. Herandez, C. Lacpatia, J.-F. Pinton, and 
G. R. Chavarria: Physica A {\bf 340} (2004) 240. 

\bibitem{gft2}
S. Sasa: nlin/0010026. 

\bibitem{uchida}
K. S. Uchida and S. Yumura:  J.  Cell Sci. {\bf 117} (2004) 1443.  

\bibitem{eff}
In non-equilibrium systems, an effective temperature is often higher 
than the temperature of the environment. See A. Crisanti and F. Ritort:  
J. Phys. A {\bf 36} (2003) R181; K. Hayashi and M. Takano: Biophys. J. 
{\bf 93} (2007) 895.  

\bibitem{miyoshi}
H. Miyoshi, N. Masaki, and Y. Tsuchiya: Protoplasma {\bf 222} (2003) 175.  

\bibitem{takagi}
H. Takagi, M. J. Sato, T. Yanagida, and M. Ueda: submitted to Phys. Rev. 
Lett. 

\bibitem{Flyvbjerg}
D. Selmeczi, S. Mosler, P. H. Hagedorn, N. B. Larsen, and H. Flyvbjerg: 
Biophys. J. {\bf 89} (2005) 912. 

\bibitem{kurchan}
F. Zamponi, F. Bonetto, L. F. Cugliandolo, and J. Kurchan: J. Stat. Mech. 
(2005) P09013. 


\end{thebibliography}
\end{document}